\documentclass[journal=jctcce,manuscript=article]{achemso}
\usepackage[utf8]{inputenc}

\usepackage{amsmath,amsfonts,amssymb}
\usepackage{mathtools}
\usepackage[nice]{nicefrac}
\usepackage{xfrac}
\usepackage{pdfpages}

\usepackage{booktabs}
\usepackage{siunitx}[=v2]
\DeclareSIUnit{\calorie}{cal}
\usepackage[vlined,ruled]{algorithm2e}
\DeclareMathOperator{\Id}{Id}
\DeclareMathOperator{\Tr}{Tr}
\DeclareMathOperator{\ExpP}{Exp}
\DeclareMathOperator{\LogP}{Log}
\newcommand{\Exp}{\ExpP_{D_0}}
\newcommand{\Log}{\LogP_{D_0}}
\DeclarePairedDelimiterX\norm[1]{\lVert}{\rVert}{#1}
\DeclarePairedDelimiterX\set[1]{\lbrace}{\rbrace}{
  \newcommand{\given}{
    \mathchoice{\:}{\:}{\,}{\,}
    \delimsize\vert
    \allowbreak
    \mathchoice{\:}{\:}{\,}{\,}
    \mathopen{}
  }#1
}
\newcommand{\R}{\mathbb{R}}
\newcommand{\Tg}{\mathcal{T}}
\newcommand{\manif}[1]{\mathcal{#1}}
\newcommand{\Gr}{\manif{G}r}
\newcommand{\St}{\manif{S}t}
\newcommand{\tr}[1]{ #1^{\mathsf{T}} }

\newcommand{\lang}[1]{\texttt{\textsc{#1}}}
\newcommand{\bp}{{\mathsf R}}
\newcommand{\Nt}{{N_t}}

\usepackage{relsize}
\usepackage{etoolbox}
\newcommand{\Gext}[1]{%
  \mbox{%
    \textsc{g-e}{\smaller[1]xt}%
    \ifstrempty{#1}
     {}
     {{\smaller[1](#1)}}%
   }%
}
\newcommand{\mw}{\textsc{xlbo{\smaller[1]/}mw}}

\newcommand{\newabbrev}[3]{%
  \global\newtoggle{#1}%
  \expandafter\def\csname #1-short\endcsname{\textsc{#2}}
  \expandafter\def\csname #1\endcsname{#3}
}
\newcommand{\newabbrevb}[3]{%
  \global\newtoggle{#1}%
  \expandafter\def\csname #1-short\endcsname{#2}
  \expandafter\def\csname #1\endcsname{#3}
}
\newcommand{\abs}[1]{%
  \expandafter\csname #1-short\endcsname%
}
\DeclareRobustCommand{\ab}[1]{%
  \nottoggle{#1}{\expandafter\csname #1\endcsname~(\expandafter\csname #1-short\endcsname)}{\expandafter\csname #1-short\endcsname}%
  \global\toggletrue{#1}%
}
\DeclareRobustCommand{\abpl}[1]{%
  \nottoggle{#1}{\expandafter\csname #1\endcsname s~(\expandafter\csname #1-short\endcsname)}{\expandafter\csname #1-short\endcsname}%
  \global\toggletrue{#1}%
}
\newabbrev{ace}{ace}{atomic cluster expansion}
\newabbrev{soap}{soap}{smooth overlap of atomic positions}
\newabbrev{admp}{admp}{atom-centered density matrix propagation}
\newabbrev{bomd}{bomd}{Born--Oppenheimer molecular dynamics}
\newabbrev{cpmd}{cpmd}{Car--Parrinello extended Lagrangian method}
\newabbrev{dft}{dft}{density functional theory}
\newabbrev{hf}{hf}{Hartree--Fock}
\newabbrev{ks}{ks}{Kohn--Sham}
\newabbrev{ltd}{ltd}{long-time drift}
\newabbrev{md}{md}{molecular dynamics}
\newabbrev{ml}{ml}{machine-learning}
\newabbrev{mm}{mm}{molecular mechanics}
\newabbrev{mo}{mo}{molecular orbital}
\newabbrev{pes}{pes}{potential energy surface}
\newabbrev{qm}{qm}{quantum mechanics}
\newabbrev{rms}{rms}{root-mean-square}
\newabbrev{scf}{scf}{self-consistent field}
\newabbrev{stf}{stf}{short-time fluctuation}
\newabbrev{svd}{svd}{singular value decomposition}
\newabbrev{xlbo}{xlbo}{extended Lagrangian Born--Oppenheimer}
\newabbrev{appa}{a{\smaller[1]pp}a}{acid phosphatase}
\newabbrev{dmabn}{dmabn}{dimethylaminobenzonitrile}
\newabbrev{ocp}{ocp}{orange carotenoid protein}
\newabbrev{thf}{{\smaller[1]3}hf}{3-hydroxyflavone}
\newabbrevb{gext}{\Gext{}}{Grassmann extrapolation}

\title{Grassmann extrapolation of density matrices for Born-Oppenheimer molecular dynamics}

\author{\'Etienne~Polack}
\affiliation{Laboratoire de Math\'ematiques de Besan\c{c}on,
  UMR CNRS 6623,
  Universit\'e Bourgogne Franche-Comt\'e,
  16~route de~Gray,
  25030~Besan\c{c}on,
  France}
\author{Genevi\`eve~Dusson}
\affiliation{Laboratoire de Math\'ematiques de Besan\c{c}on,
  UMR CNRS 6623,
  Universit\'e Bourgogne Franche-Comt\'e,
  16~route de~Gray,
  25030~Besan\c{c}on,
  France}
  \author{Benjamin~Stamm}
\affiliation{Department of Mathematics,
  RWTH Aachen University,
  Schinkelstr.\@~2,
  52062~Aachen,
  Germany}
\author{Filippo~Lipparini}
\affiliation{Dipartimento di Chimica e Chimica Industriale,
  Univerist\`a di Pisa,
  Via G.\@~Moruzzi 13,
  I-56124~Pisa,
  Italy}
\email{filippo.lipparini@unipi.it}

\begin{document}

\maketitle

\begin{abstract}
Born--Oppenheimer Molecular Dynamics (\abs{bomd}) is a powerful but expensive technique. The main bottleneck in a density functional theory \abs{bomd} calculation is the solution to the Kohn--Sham (\abs{ks}) equations, that requires an iterative procedure that starts from a guess for the density matrix. Converged densities from previous points in the trajectory can be used to extrapolate a new guess, however, the non-linear constraint that an idempotent density needs to satisfy make the direct use of standard linear extrapolation techniques not possible. In this contribution, we introduce a locally bijective map between the manifold where the density is defined and its tangent space, so that linear extrapolation can be performed in a vector space while, at the same time, retaining the correct physical properties of the extrapolated density using molecular descriptors. We apply the method to real-life, multiscale polarizable \abs{qm}/\abs{mm} \abs{bomd} simulations, showing that sizeable performance gains can be achieved, especially when a tighter convergence to the \abs{ks} equations is required.
\end{abstract}

\section{Introduction}

Ab-initio \ab{bomd} is one of the most powerful and versatile techniques in computational
chemistry, but its computational cost represents a big limitation to its routine use in
quantum chemistry.
To perform a \ab{bomd} simulation, one needs to solve the \ab{qm} equations, usually \ab{ks}
\ab{dft}, at each step, before computing the forces and propagating the trajectory of the
nuclei. The iterative \ab{scf} procedure is expensive, as it requires to build at each
iteration the \ab{ks} matrix and to diagonalize it. Convergence can require tens of
iterations, making the overall procedure, which has to be repeated a very large number of
times, very expensive.
To reduce the cost of \ab{bomd} simulations, it is therefore paramount to be able to perform
as little iterations as possible while, at the same time, obtaining an \ab{scf} solution
accurate enough to afford stable dynamics.

From a conceptual point of view, at each step of a \ab{bomd} simulation, a map is built from
the molecular geometry to the \ab{scf} density, and then to the energy and forces.
The former map, in practice, requires the solution to the \ab{scf} problem and is not only
very complex, but also highly non-linear.
However, the propagation of the \ab{md} trajectory uses short, finite time steps, so that
the converged densities at previous steps, and thus at similar geometries, are available.
As a consequence, the geometry to density map can be in principle approximated by
extrapolating the available densities at previous steps.
The formulation of effective extrapolation schemes has been the object of several previous
works~\cite{fang_existence_2016}.
Among the proposed strategies, one for density matrix extrapolation was developed by
\citet{alfe_ab_1999}, as a generalization of the wavefunction extrapolation method by
\citet{arias_ab_1992}, which is based on a least-squares regression on a few previous atomic
positions.
The main difficulty in performing an extrapolation of the density matrix stems from the
non-linearity of the problem.
In other words, a linear combination of idempotent density matrices is not an idempotent
density matrix, as density matrices are elements of a manifold and not of a vector space.
To circumvent this problem, strategies that extrapolate the Fock or \ab{ks}
matrix~\cite{pulay_fock_2004,herbert_accelerated_2005} or that use orbital transformation
methods~\cite{hutter_exponential_1994,vandevondele_efficient_2003,vandevondele_quickstep_2005}
have been proposed.

A completely different strategy has been proposed by Niklasson and
coworkers~\cite{Niklasson_PRL_TRBOMD,niklasson_extended_2008,niklasson_extended_2009}.
In the \ab{xlbo} method, an auxiliary density is propagated in a time-reversible fashion and
then used as a guess for the \ab{scf} procedure.
The strategy is particularly successful, as it combines an accurate guess with excellent
stability properties.
In particular, the \ab{xlbo} method allows one to perform accurate simulations converging
the \ab{scf} to average values (for instance, \num{e-5} in the \ab{rms} norm of the density
increment), which are usually insufficient to compute accurate forces.
An \ab{xlbo}-based \ab{bomd} strategy has been recently developed by some of us in the
context of polarizable multiscale \ab{bomd} simulations of both ground and excited
states~\cite{Loco2017,Loco_CS_PB,Nottoli2020,Bondanza_PCCP_PerspectivePol}.
Multiscale strategies can be efficiently combined, in a focused model spirit, to \ab{bomd}
simulations to extend the size of treatable systems. Using a polarizable embedding allows
one to achieve good accuracy in the description of environmental effects, especially if
excited states or molecular properties are to be computed. In such a context, the \ab{xlbo}
guessing strategy allows one to perform stable simulation even using the modes \num{e-5}
\ab{rms} convergence threshold, which, thanks to the quality of the \ab{xlbo} guess,
typically requires only about \num{4} \ab{scf} iterations.
Recently, \ab{scf}-less formulations of the \ab{xlbo} schemes have also been
proposed.\cite{Niklasson_JCP_NGXLBO,Niklasson_JCTC_NGXLBO}

Unfortunately, the performances of the \ab{xlbo}-based \ab{bomd} scheme are not so good when
a tighter \ab{scf} convergence is required, which can be the case when one wants to perform
\ab{md} simulations using post \ab{hf} methods or for excited states described in a
time-dependent \ab{dft} framework~\cite{Nottoli2020,Nottoli_JCP_2021}.
In fact, such methods require the solution to a second set of \ab{qm} equations which are
typically non-variational, making them more susceptible to numerical errors and
instabilities.
Computing the forces for non-\ab{scf} energies requires therefore a more accurate \ab{scf}
solution.

The present work builds on all previous methods for density matrix extrapolation and aims at
proposing a simple framework to overcome the difficulties associated with the non-linearity
of the problem.
The strategy that we propose is based on a differential geometry approach and is
particularly simple.
First, we introduce a molecular descriptor, \latin{i.e.}, a function of the molecular
geometry and other molecular parameters that represents the molecular structure in a natural
way that respects the invariance properties of the molecule within a vector space.
At the $(n+1)$-th step of an \ab{md} trajectory, we fit the new descriptor in a least-square
fashion using the descriptors available at a number of previous steps and obtain a new set
of coefficients.
However, we do not use them to directly extrapolate the density.
Instead, we first map the unknown density matrix, that we aim to approximate, from the
manifold where it is defined to its tangent space.
We then perform the extrapolation to approximate the representative density matrix in the
tangent space, before mapping this approximation back to the manifold in order to obtain an
extrapolated density matrix that satisfies the required physical constraints.
This geometrical strategy, that has recently been introduced in the context of density
matrix approximation by us~\cite{polack_approximation_2020}, allows one to use standard
linear extrapolation machinery without worrying about the non-linear physical constraints on
the density matrix, since both the space of descriptors and the tangent space are vector
spaces.
As the mapping between the manifold and the tangent space is locally bijective, no concerns
about redundant degrees of freedom (such as rotations that mix occupied orbitals) arise.
The map and its inverse, which are known as Grassmann Logarithm and Exponential, are easily
computed and the implementation of the strategy is straightforward.
We shall denote this approach as \ab{gext}.

In this contribution, we choose a simple, yet effective molecular descriptor and, for the
extrapolation, a least square strategy.
These are not the only choices.
As our strategy allows one to use any linear extrapolation technique between two vector
spaces, which can be in turn coupled with any choice of molecular descriptor, more advanced
strategies can be proposed, including machine learning.
Our approach ensures that the extrapolated density, independent of how it is obtained,
satisfies all the physical requirements of a density stemming from a single Slater
determinant.

The paper is organized as follows.
In the upcoming Section~\ref{sec:theo}, we present all necessary theoretical foundations
required for the development and implementation of the presented \ab{gext} approach.
Section~\ref{sec:num-tests} then presents detailed numerical tests illustrating the
performance of the extrapolation scheme, including realistic applications of \ab{bomd}
within a \ab{qm}/\ab{mm}-context before we draw the conclusion in
Section~\ref{sec:conclusion}.

\section{Theory}\label{sec:theo}

We consider Born--Oppenheimer ab-initio \ab{bomd} simulations where the position vector
$\bp \in \R^{3M}$ evolves in time according to classical mechanics as
\begin{equation}
  M_i \, \ddot{\bp}_i(t) =   F_i(t, \bp(t)),
\end{equation}
where ${\bp}_i(t),F_i(t)\in\R^3$ denote the position of the $i$-th atom with mass
$M_i$ respectively the force acting on it at time $t$.
We consider a general \ab{qm}/\ab{mm}-method but the setting also trivially applies to
pure \ab{qm}-models.
The forces at a given time $t$ and position $\bp$ of the nuclei arise from different
interactions, namely \ab{qm}-\ab{qm}, \ab{qm}-\ab{mm} and \ab{mm}-\ab{mm}
interactions.
The computationally expensive part is to determine the state of the electronic structure,
which is modelled here at the \ab{dft} level with a given basis set of dimension
$\mathcal N$. Note that considering \ab{hf} instead of \ab{dft} would not change much in
the presentation of the method.
It consists of computing the instantaneous non-linear eigenvalue problem
\begin{equation}
    \label{eq:EVP}
  \left\{ \begin{aligned}
    & F_{\bp}(D_{\bp})C_{\bp} = S_{\bp}C_{\bp}E_{\bp} \\
    & \tr{C_{\bp}} S_{\bp} C_{\bp} = \Id_{N} \\
    & D_{\bp} = C_{\bp}\tr{C_{\bp}}
  \end{aligned}\right.,
\end{equation}
where $C_{\bp}\in \R^{\mathcal N \times N}$ and
$D_{\bp}\in \R^{\mathcal N \times \mathcal N}$ denote the coefficients respectively of the
occupied orbitals and density matrix and $E_{\bp}\in \R^{\mathcal N \times N}$ the diagonal
matrix containing the energy levels.
Further, $F_{\bp}$ denotes the \ab{dft}-operator acting on the density matrix and $S_{\bp}$
the customary overlap matrix.

At this point it is useful to note that the slightly modified coefficient matrix
$\widetilde {C}_{\bp} \coloneqq S_{\bp}^{\nicefrac{1}{2}}{C}_{\bp} $ belongs to the
so-called Stiefel manifold defined as follows
\begin{equation}
  \St(N, \mathcal N) \coloneqq \set*{V \in \R^{\mathcal N\times  N} \given \tr{V}V = \Id_{N}},
\end{equation}
due to the second equation in Equation~\eqref{eq:EVP}.
In consequence the normalized density matrix
$\widetilde D_{\bp} = \widetilde C_{\bp}\tr{\widetilde C_{\bp}} = S_{\bp}^{\nicefrac{1}{2}} {D}_{\bp} S_{\bp}^{\nicefrac{1}{2}} $
belongs to the following set
\begin{equation}
  \Gr(N, \mathcal{N}) \coloneqq \set*{D \in \R^{\mathcal{N}\times \mathcal{N}} \given D^{2} = D, \; \tr{D} = D, \; \Tr D = N},
\end{equation}
which can be identified with the Grassmann manifold of $N$-dimensional subspaces of
$\R^{\mathcal N}$ by means of the spectral projectors.
In the following, we always assume that the density matrix has been orthonormalized, and
therefore drop the $\sim$ from the notation.
For any $D\in \Gr(N, \mathcal{N})$, one can associate the tangent space $\mathcal T_D$ which
has the structure of a vector space.
The evolution of the electronic structure can therefore be seen as a trajectory
$t \mapsto D_{\bp(t)}$ on $\Gr(N, \mathcal{N})$ where $t \mapsto \bp(t)$ denotes the
trajectory of the nuclei.

The goal of the present work is to find a good approximation for the electronic density
matrix at the next step of \ab{md} trajectory by extrapolating the densities at previous
steps.
More precisely, based on the knowledge of the density matrices $D_i \coloneqq D_{\bp(t_i)}$,
$i=n-N_{t},\ldots,n-1$, at $N_{t}$ previous times $t_i$, one aims to compute an accurate
guess of the density matrix $D_n$ at time $t_n$.

Thus, the problem formulation can be seen as an extrapolation problem of the following form:
given the set of couples $(\bp(t_i),D_i)$ and a new position vector $\bp(t_n)$, provide a
guess for the solution $D_n$.
Here and in the remaining part of the article, we restrict ourselves on the positions of the
\ab{qm}-atoms, \latin{i.e.}, with slight abuse of notation we denote from now on by $\bp$
the set of \ab{qm}-positions only, even within a \ab{qm}/\ab{mm}-context.

In order to approximate the mapping $\bp \mapsto D_{\bp}$, we split this mapping in several
sub-maps that will be composed as follows:
\begin{equation}
    \label{eq:all_maps}
  \begin{array}{@{}r@{\,}c@{\,}c@{\,}c@{\,}c@{\,}c@{\,}l@{}}
    \R^{3M}& \to &\mathcal M& \to &\mathcal T_{D_0}& \to &\Gr(N,\mathcal N)
    \\
    \bp& \mapsto &d_\bp& \mapsto &\Gamma_\bp& \mapsto &D_{\bp} = \Exp(\Gamma_\bp),
  \end{array}
  \end{equation} where the first line shows the concatenation of maps in terms of spaces and
the second in terms of variables.
The different mappings will be presented and motivated in the following.

The first map is a mapping of the nuclear coordinates $\bp \in \R^{3M}$ to a (possibly
high-dimensional) molecular descriptor $d_\bp \in \mathcal{M}$ that accounts for certain
symmetries and invariances of the molecule.
The last map, known as the Grassmann exponential, is introduced in order to obtain a
resulting density matrix belonging to $\Gr(N,\mathcal N)$ and thus to guarantee that the
guess fulfils all properties of a density matrix. As $\Gr(N,\mathcal N)$ is a manifold this
is not straightforward.
The second mapping is the one that we aim to approximate but before we do that, let us first
introduce those two special mappings, \latin{i.e.}, the molecular descriptor and the
Grassmann exponential, in more details.

\subsection{Molecular descriptors}\label{sec:descriptors}

The map $\bp \mapsto d_\bp$ is a map from atomic positions to molecular descriptors.
These descriptors are used as fingerprints for the considered molecular configurations.
Such molecular descriptors have been widely used in the past decades \latin{e.g.}, to learn
\abpl{pes}~\cite{Behler2011, Rupp2012, Bartok2013, Behler2015, Manzhos2015, Chmiela:2017ff,
Chmiela2018}, or to predict other quantities of interest.
Among widely used descriptors, one can find Behler--Parinello symmetry
functions~\cite{Behler2007-pg}, Coulomb matrix~\cite{Rupp2012-hu},
\ab{soap}~\cite{Goscinski2020-jp}, permutationally invariant
polynomials~\cite{Braams2009-wi}, or the \ab{ace}~\cite{Drautz2019-rb,Bachmayr2019-ec}.
These molecular descriptors are usually designed to retain similar symmetries as the
targeted quantities of interest.

In this work, the quantity we are approximating is the density matrix, which is invariant
with respect to translations as well as permutations of like particles.
The transformation of the density matrix with respect to a global rotation of the system
depends on the implementation, as it is possible to consider either a fixed Cartesian frame
or one that moves with respect to the molecular system. In the former
case, there is an equivariance with respect to rotations of the molecular system,
while in the latter, the density matrix is invariant. We should therefore in principle use a
molecular descriptor satisfying those properties.

However, the symmetry properties we will rely on are mostly translation and rotation
invariance.
Therefore, we will use a simple descriptor in form of the Coulomb-matrix denoted by $d_\bp$,
given by
\begin{equation}
  (d_\bp)_{ij} =
  \begin{dcases}
    0.5 z_i^{2.4} & \text{ if  \( i = j \) } \\
    \frac{z_i z_j}{\norm{\bp(t_i) - \bp(t_j)}} & \text{ otherwise }
  \end{dcases}.
\end{equation}
Note that such a descriptor is not invariant (nor equivariant) with respect to permutations
of identical particles.
However, we have found this descriptor to offer a good trade-off between simplicity and
efficiency.
Note that since we aim to extrapolate the density matrix from previous time-steps,
permutations of identical particles never occur from one time-step to another and we do not
need to rely on this property.
Nevertheless, we expect that a better description could be achieved by using more flexible
descriptors, such as \ab{ace} polynomials or the \ab{soap} descriptors, where the
descriptors themselves can be tuned.

\subsection{The Grassmann exponential}

We only give a brief overview as the technical details have already been reported
elsewhere~\cite{edelman_geometry_1998,zimmermann_manifold_2019,polack_approximation_2020}.
The set $\Gr(N,\mathcal N)$ is a smooth manifold and thus, at any point, say
$D_0\in\Gr(N,\mathcal N)$ in our application, there exists the tangent space
$\mathcal T_{D_0}$ such that one can associate nearby points $D\in\Gr(N,\mathcal N)$ to
tangent vectors $\Gamma(D) \in \mathcal T_{D_0}$. The mapping $D\mapsto \Log(D)= \Gamma(D)$
is known as the Grassmann logarithm and its inverse mapping as the Grassmann exponential
$\Gamma \mapsto \Exp(\Gamma)=D$.
There also holds that $\Log(D_0)=0$ and $\Exp(0)=D_0$.
These mappings are not only abstract tools from differential geometry but can be computed by
means of performing an
\ab{svd}~\cite{edelman_geometry_1998,zimmermann_manifold_2019,polack_approximation_2020}.
In our application we use the same reference point $D_0$ in all cases which brings some
computational advantages as will be discussed in more detail the the upcoming
Section~\ref{ssec:ApproxPb}.

\subsection{The approximation problem}\label{ssec:ApproxPb}

Since the tangent space~\( \Tg_{D_0} \) is a (linear) vector space, we can now aim to
approximate the mapped density matrix on the tangent space~\( \Tg_{D_{0}} \).
To simplify the presentation, we shift the indices in the following and describe the
extrapolation method for the first $N_{t}$ time steps. In the general setting, we should
consider the positions $\bp(t_{i})$ for $i=n-N_{t}, \ldots, n-1$, to extrapolate the density
matrix at position $\bp(t_{n})$, where $n$ is the current time step of the \ab{md}.
We look for parameter functions~\( c_{i} \), such that, given previous
snapshots~\( \Gamma_{i} = \Log(D_i) \) for~\( i \) from \( 1 \)~to~\( N_{t} \),
corresponding to some~\( \bp(t_i) \)'s, the approximation of any density matrix on the
tangent space is written as
\begin{equation}
    \label{eq:gamma_app}
    \bp \mapsto
    \Gamma_{\mathrm{app}}(\bp) = \sum_{i=1}^{\Nt} c_{\bp,i} \, \Gamma_i \in \Tg_{D_0},
\end{equation}
with $\Gamma_i = \Gamma_{\bp(t_i)}$.

The question is then how to find these coefficient functions~\( c_{\bp,i} \) and we propose
to find those via the resolution of a (standard) least-square minimization problem.
For a given position~\( \bp \), we look for coefficients that minimise
the~\( \ell^{2} \)--error between the descriptor~\( d_\bp \) and a linear combination of the
previous ones $d_{\bp(t_i)}$
\begin{equation}
  \label{eq:LSQ_pb}
  \min_{c_{\bp} \in \R^{N_{t}}}
  \;\;
  \norm*{ d_\bp - \sum_{i=1}^{N_{t}} c_{\bp,i} d_{\bp(t_i)} }^{2}.
\end{equation}

In matrix form, this simply reads
\begin{equation}
  \min_{c_{\bp} \in \R^{N_{t}}}  \; \; \norm*{ d_\bp - \tr{P}  c_\bp }^{2},
\end{equation}
where~\( P \) is the matrix of size~\( N_{t} \times N_{d} \) containing the descriptors
$P_{i,j} \coloneqq (d_{\bp(t_i)})_{j}$.
Note that we only fit on the level of the descriptor, \latin{i.e.}, the mapping from the
position vector $\bp$ to the descriptor $d_\bp$, and that this method is similar to the ones
used by \citet{arias_ab_1992, alfe_ab_1999}, where the descriptors they used were the
positions of the atoms and only considered the previous three time-steps of the \ab{md}.

If the system is underdetermined, we select the vector~\( c_\bp \) that has the smallest
norm.
However, in general, the system is overdetermined as we have more descriptors than
snapshots.
This implies that this formulation verifies the interpolation principle: for every~\( i \)
and~\( j \) from \( 1 \)~to~\( N_{t} \), the solution of Problem~\eqref{eq:LSQ_pb} at the
positions $\bp(t_j)$ satisfies~\( c_{\bp(t_j),i} = \delta_{ji} \).

In principle, should we consider a large amount of previous descriptors, then the system may
become undetermined and violates the interpolation principle.
To mitigate this, we can use a stabilization scheme, as explained in the upcoming
subsection.

Note that once we have computed the coefficients $c_{\bp}$ by solving
Problem~\eqref{pb_reg}, one computes the initial guess for the density by using the same
coefficients in the linear combination on the tangent space as in
Equation~\eqref{eq:gamma_app} and finally take the exponential (see
Equation~\eqref{eq:all_maps}).
The rational for this step is that, if the second mapping in Equation~\eqref{eq:all_maps},
that we denote here by $\mathcal F \colon \mathcal M \to \mathcal T_{D_0}$, was linear, then
there would hold
\begin{equation}
  \label{eq:linearity}
  \mathcal F\left(\sum_{i=1}^{N_t} c_{\bp,i} d_{\bp_{i}}\right)
  = \sum_{i=1}^{N_t} c_{\bp,i}  \mathcal F(d_{\bp_{i}}) = \sum_{i=1}^{N_t} c_{\bp,i} \Gamma_i.
\end{equation}
In practice, the mapping is however not linear and this approach works well in the test
cases we considered.
A possible explanation for this is the unfolding of the nuclear coordinates into a
high-dimensional descriptor-space $\mathcal M$.
Indeed, the high-dimensionality of $\mathcal M$ seems to allow an accurate approximation of
$\mathcal F$ by a linear map.
Further, if the system is overdetermined, the scheme satisfies the interpolation property
\( \Gamma_{j} = \Gamma(\bp(t_j)) \), and hence we recover the expected density
matrix~\( D_{\bp(t_{j})} = \Exp(\Gamma_j) \).

\subsubsection{Stabilization}

To stabilize the extrapolation by limiting high oscillations of the coefficients, we apply a
Tikhonov regularization
\begin{equation}
  \min_{c_{\bp} \in \R^{N_{t}}}  \left (
    \norm*{ d_\bp - \sum_{i=1}^{N_{t}} c_{\bp,i} d_{\bp_{i}} }^{2}
    + \varepsilon \, \norm*{c_{\bp}}^{2}
  \right),
\end{equation}
for some choice of~\( \varepsilon \).
This problem is always well-posed, and corresponds to solving the following problem
\begin{equation}
\label{pb_reg}
  \min_{c_{\bp} \in \R^{N_{t}}} \; \; \norm*{ \widetilde{d_\bp} - \tr{\widetilde{P}} \cdot c_\bp }^{2},
\end{equation}
where~\( \widetilde{d_\bp} \in \R^{N_{d} + N_{t}} \) is the vector~\( d_\bp \) padded
with~\( N_{t} \) zeros and~\( \widetilde{P} \in \R^{N_{t}} \times \R^{N_{d} + N_{t}} \) is
the~\( P \) matrix padded with the square diagonal matrix~\( \varepsilon\Id_{N_{t}} \).
We observe in practice that using such a stabilization makes possible to use more previous
points without degradation of the initial guess.

\subsection{The final algorithm}

\begin{algorithm}[!htb]
  \SetKwData{c}{c}%
  \SetKwData{desc}{desc}%
  \SetKwData{cmat}{cmat}%
  \SetKwData{gmat}{gmat}%
  \SetKwData{cref}{cref}%
  \SetKwFunction{Logf}{Log}%
  \SetKwFunction{Expf}{Exp}%
  \SetKwFunction{Orthonormalization}{Orthonormalization}%
  \SetKwFunction{Stabilization}{Stabilization}%
  \SetKwFunction{LeastSquares}{LeastSquares}%

  \KwData{Array \desc containing the descriptors for~\( k \) previous time-steps,
    \( p_{n} \) the descriptor for the current position,
    \( C_{n-1} \) and \( S_{n-1} \) respectively the molecular orbitals and overlap matrices
    of the previous time-step,
    and \cref the reference point on the Grassmannian
  }
  \KwResult{Guess density matrix for time-step~\( n > 1 \)}

  \BlankLine%

  \Begin{%
    \BlankLine%
    \cmat{:, :, $n-1$} \( \leftarrow \) \Orthonormalization{\( C_{n-1} \), \( S_{n-1} \)}\;%

    \gmat{:, :, $n-1$} \( \leftarrow \) \Logf{\cref, \cmat{:, :, $n-1$}}\;%

    \desc, \( p_{n} \) \( \leftarrow \) \Stabilization{\desc, \( p_{n} \)}\;%

    \c\ \( \leftarrow \) \LeastSquares{\desc, \( p_{n} \)}\;%

    \( \Gamma_{\mathrm{app}} \leftarrow \sum_{i=n-1-k}^{n-1} \c{i} \cdot \gmat{:, :, i} \)\;%

    \( C_{\mathrm{app}} \leftarrow \) \Expf{\cref, \( \Gamma_{\mathrm{app}} \)}\;%

    \Return\ \( 2 \cdot C_{\mathrm{app}} \cdot \tr{C_{\mathrm{app}}} \)\;%
  }%
  \caption{Density extrapolation framework \ab{gext}\label{algo1}}
\end{algorithm}

Given previous density matrices $D_{\bp(t_j)}$ for $j = 1,\ldots,\Nt$, the initial guess is
computed following Algorithm~\ref{algo1}.
That is, we start by computing the logarithms of the density matrices $D_{\bp(t_j)}$, from
the coefficients $C_{\bp(t_j)}$ that are first orthonormalized by performing
$\widetilde {C}_{\bp} = S_{\bp}^{\nicefrac{1}{2}}{C}_{\bp} $.
Here, we remark that we assume that the density matrices $D_{\bp(t_j)}$ have been previously
L\"owdin orthonormalized.

We then compute the descriptors needed to build the $\widetilde{P}$ matrix and solve
Problem~\eqref{pb_reg}.
This provides the coefficients in the linear combination of the $\Gamma_i's$ on the tangent
space.
Finally, we compute the exponential of the linear combination in order to obtain the
predicted density matrix.

Note that the reference point $D_0$ is chosen once and for all, which makes the computations
of these logarithms lighter, even though there is no theoretical justification for keeping a
single point~\( D_{0} \) as a reference.
Indeed, it is known that the formulae are only correct locally (around \( D_{0} \)) on the
manifold.
However, in practice we have never observed the need to change the reference point.
This enables us to compute only one logarithmic map per time step; and hence, only two
\abpl{svd} in total per time step.
To have a robust algorithm that will work even in this edge case, it will be sufficient to
check that the exponential and logarithmic maps are still inverse of one another.

Finally, to be on the safe-side with respect to the computations of the exponential, we have
added a check on the orthogonality of the matrix that is obtained: If the residue is higher
than a certain threshold, we then perform an orthogonalization of the result.

\section{Numerical tests}\label{sec:num-tests}

In this section we present a series of numerical tests of the newly developed strategy.
We test our method on four different systems.
All the systems have been object of a previous or current study by some of us, and can
therefore be considered representative of real-life applications.

The first system is \ab{thf} in acetonitrile~\cite{Nottoli_JCP_2021}.
Two systems (\abs{ocp} and \abs{appa}) are chromophores embedded in a biological matrix ---
namely, a carotenoid in the \ab{ocp} and flavine in \ab{appa}, a blue light-using flavine
photoreceptor~\cite{Bondanza_Chem_OCP,Bondanza_JACS_OCP,Macaluso_CS_AppA}.
The fourth system is \ab{dmabn} in methanol~\cite{Nottoli2020}.
The main characteristics of the systems used for testing are recapitulated in
Table~\ref{tab:systems}.

\begin{table}[!htb]
    \centering
    \begin{tabular}{lS[table-format=3.0]S[table-format=5.0]S[table-format=4.0]}
        \toprule
         System & $N_{QM}$ & $N_{MM}$ & $\mathcal N$  \\
         \midrule
         \ab{ocp}    &     129 &    4915 & 1038 \\
         \ab{appa}   &      31 &   16449 &  309 \\
         \ab{dmabn}  &      21 &    6843 &  185 \\
         \ab{thf}    &      28 &   15018 &  290 \\
         \bottomrule
    \end{tabular}
    \caption{Overview of the system size in terms of number of \ab{qm}-atoms ($N_{QM}$),
      number of \ab{mm}-atoms ($N_{MM}$) and the total number of (\ab{qm}) basis functions
      ($\mathcal N$).\label{tab:systems}}
\end{table}

The systems used for testing include a quite large \ab{qm} chromophore, the \ab{ocp} and
three medium-sized systems, embedded in large (\ab{appa}, \ab{thf}) and medium-sized
environments (\ab{dmabn}) and are representative of different possible scenarios.

To test the performances of the new \ab{gext} strategy, we performed three sets of short
(\SI{1}{\ps}) multiscale \ab{bomd} simulations on \ab{ocp}, \ab{appa}, \ab{thf}, and
\ab{dmabn}.
\ab{ks} density functional theory was used to model the \ab{qm} subsystem, using the
B3LYP~\cite{B3LYP} hybrid functional and Pople's 6-31G(d) basis set~\cite{Hehre1972}.
For the stability and energy conservation of the method, we did a longer and more realistic
simulations of \SI{10}{\ps} on \ab{thf}, where the flavone moiety was described using the
$\omega$B97X hybrid functional~\cite{wB97X-D} and Pople's 6-31G(d) basis set.
In all cases, the environment was modeled using the AMOEBA polarizable force
field~\cite{ponder2010}.

All the simulations have been performed using the Gaussian--Tinker interface previously
developed by some of us~\cite{Loco2017,Loco_CS_PB}.
In particular, we use a locally modified development version of Gaussian~\cite{gdv} to
compute the \ab{qm}, electrostatic and polarization energy and forces, and
Tinker~\cite{Tinker} to compute all others contributions to the \ab{qm}/\ab{mm} energy.
We implemented the \ab{gext} extrapolation scheme in Tinker, that acts as the main driver
for the \ab{md} simulation, being responsible of summing together all the various
contributions to the forces and propagating the trajectory.
At each \ab{md} step, using the GauOpen interface~\cite{gauopen}, the density matrix,
\ab{mo} coefficients, and overlap matrix produced by Gaussian are retrieved.
These are used to compute the extrapolated density as described in Section~\ref{sec:theo}.
The density is then passed back to Gaussian to be used for the next \ab{md} step.
All the simulations were carried out in the NVE ensemble, using the velocity Verlet
integrator and a \SI{0.5}{\fs} time step.
Concerning stabilization, we found that good overall results were obtained using a
parameter~\( \varepsilon \coloneqq 10^{3} \cdot r_{\mathrm{scf}} \),
where~\( r_{\mathrm{scf}} \) is the tolerance of the \ab{scf} algorithm.

\subsection{Numerical results}

To assess the performance of the \ab{gext} guess we perform \SI{1}{\ps} \ab{md} simulations
on the four systems described in Section~\ref{sec:num-tests} starting from the same exact
conditions (positions and initial velocities) and using various strategies to compute the
guess density for the \ab{scf} solver.
We compare various flavors of the \ab{gext} method with the the \ab{xlbo} extrapolation
scheme~\cite{niklasson_extended_2008}.
Here, we note that the original \ab{xlbo} method performs a propagation of an auxiliary
density matrix, which is then used as a guess.
The latter is not idempotent: to restore such a property, we perform a purification step
at the beginning of the SCF procedure using McWeeny's algorithm~\cite{McWeeny_RMP_Purification}.
In the following, we therefore compare our method, where we use \num{3} to \num{6} previous
points for the fitting and extrapolation, to both the standard \ab{xlbo} and to \ab{xlbo}
followed by purification (\mw{}).
We use an \ab{scf} convergence threshold of \num{e-5} with respect to the \ab{rms} variation
of the density.

We report in Table~\ref{tab:conv5}, for each method, the average number of \ab{scf}
iterations performed along the \ab{md} simulation together with the associated standard
deviation.
As the \ab{xlbo} strategy requires \num{8} previous points, during which a standard \ab{scf}
is performed, we discard the first points from the evaluation of the aforementioned
quantities to have a fairer comparison.

We do not report the total time required to compute the guess, as it is in all cases very
small (up to \SI{0.1}{\second} wall clock time for the largest system using the \Gext{6}
guess).
This is an important consideration, as the \ab{gext} method requires one to perform various
linear-algebra operations (in particular, thin \ab{svd}) that can in principle be expensive.
Thanks to the availability of optimized LAPACK libraries, this is in practice not a problem.

\begin{table}[!htb]
  \caption{Performances of the \ab{gext} method for different number of
    extrapolation points, compared with the \ab{xlbo} algorithm with and without McWeeny
    purification. All the results were obtained using a \num{e-5} convergence threshold for
    the root-mean-square increment of the density matrix and are derived from a \SI{1}{\ps}
    long \ab{md} simulation, using a \SI{0.5}{\fs} time step. We report the average number
    of iterations required to converge the \ab{scf}, together with the associated standard
    deviation. Note that the first \num{8} steps were discarded.\label{tab:conv5}}
  \centering
  \begin{tabular}{l*8{S[table-format=1.2]}}
    \toprule
    & \multicolumn{2}{c}{\ab{ocp}} & \multicolumn{2}{c}{\ab{dmabn}} & \multicolumn{2}{c}{\ab{appa}} & \multicolumn{2}{c}{\ab{thf}}\\
    \cmidrule(r){2-3}
    \cmidrule(r){4-5}
    \cmidrule(r){6-7}
    \cmidrule(r){8-9}
    Method        & {Average} & {$\sigma$} & {Average} & {$\sigma$} & {Average} & {$\sigma$} & {Average} & {$\sigma$} \\
    \midrule
    \ab{xlbo}     & 3.82    & 0.66  & 3.98    & 0.16  & 3.00    & 0.03    & 4.00 & 0.14 \\
    \mw{}         & 2.95    & 0.31  & 3.76    & 0.56  & 3.00    & 0.34    & 3.96 & 0.31 \\
    \Gext{3}      & 2.57    & 0.84  & 3.54    & 0.78  & 2.95    & 0.50    & 3.09 & 0.41  \\
    \Gext{4}      & 2.48    & 0.88  & 3.14    & 0.62  & 2.51    & 0.50    & 3.25 & 0.68  \\
    \Gext{5}      & 2.25    & 0.96  & 3.23    & 0.75  & 2.51    & 0.50    & 3.30 & 0.72  \\
    \Gext{6}      & 2.20    & 0.96  & 2.99    & 0.02  & 2.51    & 0.50    & 3.14 & 0.56  \\
    \bottomrule
  \end{tabular}
\end{table}

From the results in Table~\ref{tab:conv5}, we see that the \ab{gext} algorithms
systematically outperforms the \ab{xlbo} method.
It is interesting to note that the McWeeny purification step has a sizeable effect on the
performances of the \ab{xlbo} method only for the largest system, \ab{ocp}, where it results
in the gain of almost one \ab{scf} iteration on average.
On the other systems, the purification step has a smaller effect.

In all the systems we tested, the performances of the \ab{gext} method are systematically
better than in \ab{xlbo}, including with McWeeny purification.
The effectiveness of the \ab{gext} extrapolation increases when going from \num{3} to
\num{6} points, but quickly stagnates.
We have performed further tests with more than \num{6} (up to \num{20}) extrapolation
points, but never noted any further gain.

We observe a reduction in the number of iterations that goes from \num{0.5} in \ab{dmabn} to
\num{0.75} in \ab{ocp} (\num{1.62} when compared to \ab{xlbo} without McWeeny purification).
We remark that these gains, while apparently not so large, are greatly amplified during the
\ab{md} simulation, due to the large number of steps that need to be performed.

The tests performed with a \num{e-5} convergence threshold are representative of a standard,
\ab{dft} ground state \ab{bomd} simulation.
When performing a more sophisticated quantum mechanical calculation, such as a \ab{bomd} on
an excited state \ab{pes}~\cite{Nottoli_JCP_2021}, such a convergence threshold may not be
sufficient to guarantee the stability of the simulation, as the \ab{scf} solution is used to
set up the linear response equations and the numerical error can be amplified, resulting in
poorly accurate forces.

We tested the \ab{gext} algorithm in its best-performing version, the one that uses six
extrapolation points, with a tighter, \num{e-7} threshold, again for the \ab{rms} variation
of the density.
The results are reported in Table~\ref{tab:conv7}, where we compare the \Gext{6} scheme with
the \ab{xlbo} method with McWeeny purification.

The \ab{xlbo} method is based on the propagation of an auxiliary density and therefore the
accuracy of the guess it generates depends little on the accuracy of the previous \ab{scf}
densities.
As a consequence, its performances are reduced if a tighter convergence is required.
The \ab{gext} guess, on the other hand, uses previously computed densities as its building
blocks and one can expect the accuracy of the resulting guess to be linked to the
convergence threshold used during the simulation.

This is exactly what we observe.
Using a threshold of \num{e-7}, the \Gext{6} guess exhibits significantly better
performances than \ab{xlbo}, gaining, on average, from about \num{0.7} to about \num{3}
\ab{scf} iterations on the tested systems.

\begin{table}[!htb]
  \caption{Performances of the \Gext{6} method compared with the \ab{xlbo} algorithm with
    McWeeny purification. All the results were obtained using a \num{e-7} convergence
    threshold for the root-mean-square increment of the density matrix and are derived from
    a \SI{1}{\ps} long \ab{md} simulation, using a \SI{0.5}{\fs} time step. We report the
    average number of iterations required to converge the \ab{scf}, together with the
    associated standard deviation. Note that the first \num{8} steps were
    discarded.\label{tab:conv7}}
  \centering
  \begin{tabular}{l*8{S[table-format=1.2]}}
    \toprule
    & \multicolumn{2}{c}{\ab{ocp}} & \multicolumn{2}{c}{\ab{dmabn}} & \multicolumn{2}{c}{\ab{appa}}  &
    \multicolumn{2}{c}{\ab{thf}}\\
    \cmidrule(r){2-3}
    \cmidrule(r){4-5}
    \cmidrule(r){6-7}
    \cmidrule(r){8-9}
    Method        & {Average} & {$\sigma$} & {Average} & {$\sigma$} & {Average} & {$\sigma$} & {Average} & {$\sigma$} \\
    \midrule
    \mw{}         & 5.02    & 0.17  & 7.30  & 0.64  & 7.49  & 0.84 & 7.47 & 0.63 \\
    \Gext{6}      & 3.58    & 0.79  & 4.23  & 0.50  & 4.39  & 0.57 & 6.81 & 0.78 \\
    \bottomrule
  \end{tabular}
\end{table}

\subsubsection{Stability}

The good performances of the \ab{gext} guess come, however, at a price, namely, the lack of
time reversibility.
We can thus expect the total energy in a NVE simulation to exhibit a \ab{ltd}.
Time reversibility and long-time energy conservation are, on the other hand, one of the
biggest strengths of the \ab{xlbo} method.

To investigate the stability of \ab{bomd} simulations using the \ab{gext} guess, we build a
challenging case, where we start a \ab{bomd} simulation far from well-equilibrated
conditions.
We use the \ab{thf} system as a test case and achieve the noisy starting conditions by
starting from a well-equilibrated structure and changing the \ab{dft} functional from B3LYP
to $\omega$B97XD\@.
This way, we have a physically acceptable structure, with no close atoms or other
problematic structural situations, but obtain starting conditions that are far from
equilibrium.

\begin{figure}[!htb]
    \centering
    \includegraphics[width=0.475\linewidth]{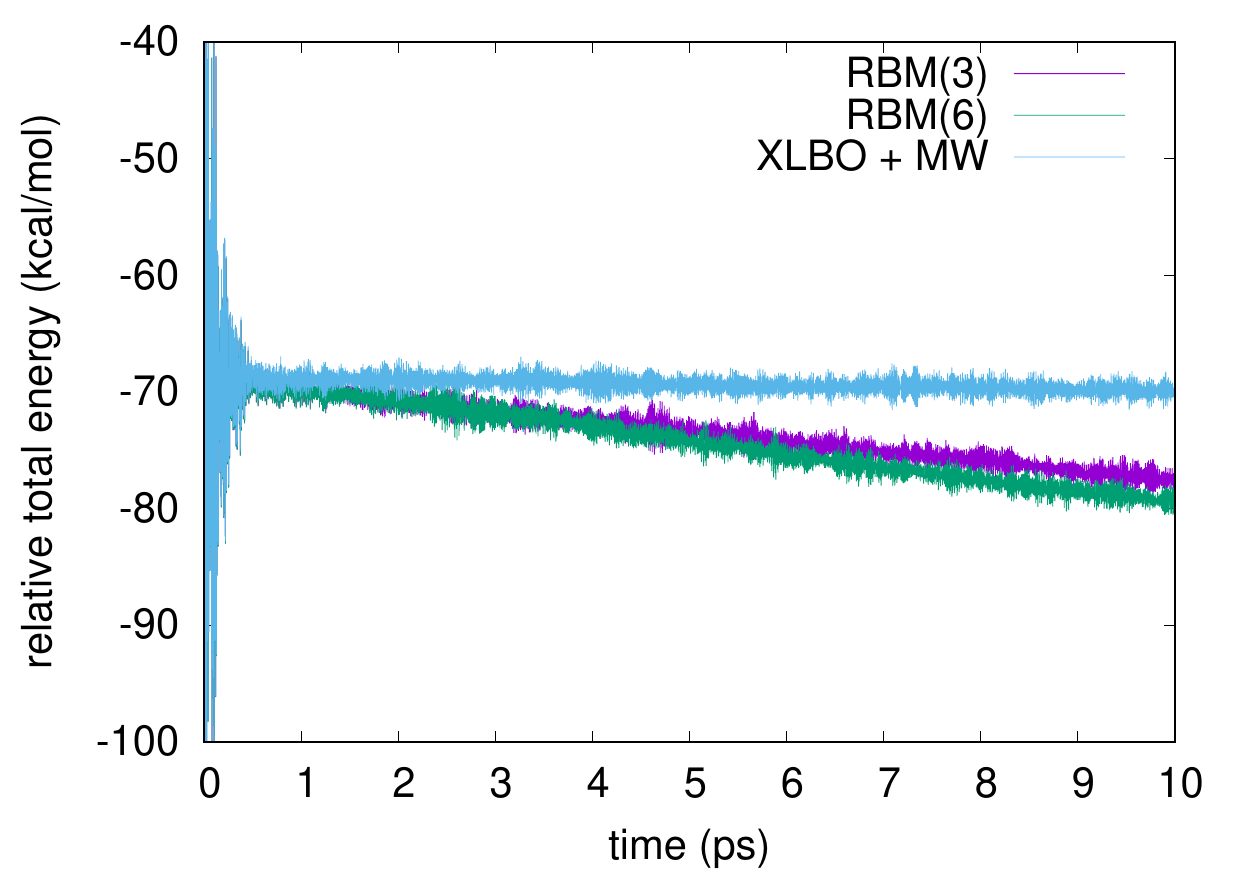}
    \hfill
    \includegraphics[width=0.475\linewidth]{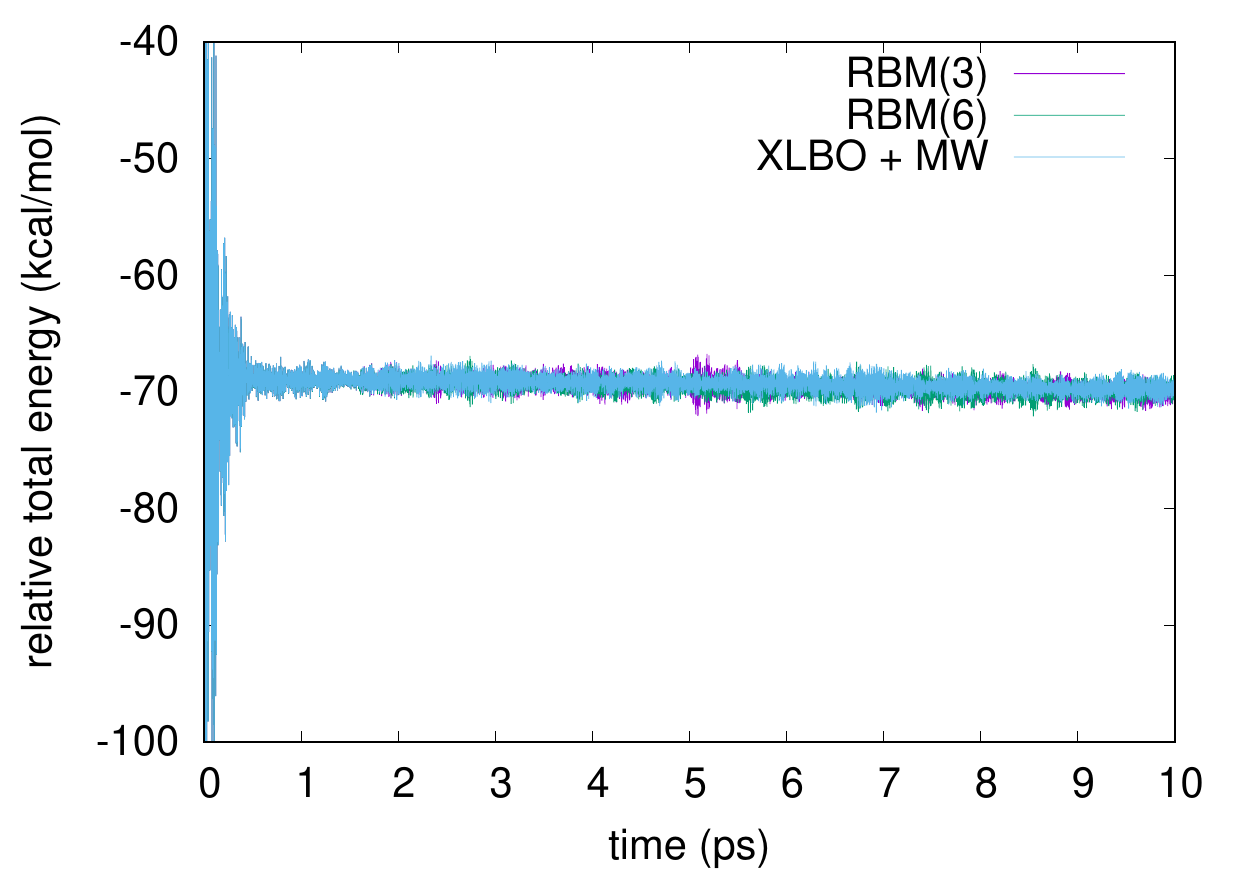}
    \caption{Total energy (\si{\kilo\calorie\per\mol}) as a function of simulation time
      (\si{\fs}) for \ab{thf} comparing \Gext{3}, \Gext{6} and \ab{xlbo} with McWeeny
      purification, using a convergence threshold for the \ab{scf} algorithm of \num{e-5}
      (left panel) and \num{e-7} (right panel). The total energy was shifted of
      \SI[retain-explicit-plus]{+505 000}{\kilo\calorie\per\mole} for
      readability.\label{fig:econs}}
\end{figure}

We report in Figure~\ref{fig:econs} the total energy along a \SI{10}{\ps} \ab{bomd}
simulation of \ab{thf} in acetonitrile using either a \num{e-5} \ab{scf} convergence
threshold (left panel) or a \num{e-7} one (right panel).
The same results for a \num{e-6} threshold are reported in the supporting information.
We compare the \Gext{3} and \Gext{6} methods to the \ab{xlbo} one including McWeeny
purification.
As already noted, while in principle the purification may spoil the time reversibility, this
has no noticeable effect in practice.

The very noisy starting conditions are apparent from the energy profiles, that exhibits
large oscillations in the first couple hundreds femtoseconds.

To better estimate the short- and long-time energy stability, we report in
Table~\ref{tab:t3} the average \ab{stf} and \ab{ltd} of the energy.
The former is computed by taking the \ab{rms} of the energy fluctuation every \SI{50}{\fs}
and averaging the results over the trajectory, discarding the first \SI{500}{\fs}, the
latter by fitting the energy with a linear function and taking the slope.

All methods show comparable short-term stability, which is to be mainly ascribed to the
chosen integration time-step.
On the other hand, from both the results in Table~\ref{tab:t3} and Figure~\ref{fig:econs},
we observe a clear drift of the energy when the \ab{gext} method is used.
In particular, the system cools of about \SI{10}{\kilo\calorie\per\mol} with either \Gext{3}
or \Gext{6}.
The \ab{xlbo} trajectory, despite the McWeeny purification, exhibits an almost perfect
energy conservation.

These results are not surprising, but should be taken into account when choosing to use the
\ab{gext} guess, which, if coupled to a \num{e-5} \ab{scf} convergence threshold, cannot
guarantee long-term energy conservation.
The drift is overall not too large and can be handled by using a thermostat.
Whether or not the trade between performances and energy conservation is acceptable for a
production simulation is a decision that ultimately lies with the user.

Increasing the accuracy of the \ab{scf} computation improves the overall stability for
\ab{gext}, which is already good at \num{e-6} and becomes virtually identical to the one
offered by the \ab{xlbo} method at \num{e-7}.

\begin{table}[!htb]
  \caption{Short and long-term stability analysis of the \Gext{3} and \Gext{6} methods,
    compared to the \ab{xlbo} algorithm with McWeeny purification, for the \ab{thf} system.
    For each method we report the \ab{stf} and the \ab{ltd} and the average number of
    \ab{scf} iterations, for three convergence thresholds of the \ab{scf}
    algorithm.\label{tab:t3}}
  \centering
  \begin{tabular}{l*3{S[table-format=1.2]S[table-format=+1.2]}}
    \toprule
              & \multicolumn{2}{c}{Conv. \num{e-5}} & \multicolumn{2}{c}{Conv. \num{e-6}} & \multicolumn{2}{c}{Conv. \num{e-7}}\\
    \cmidrule(r){2-3}
    \cmidrule(r){4-5}
    \cmidrule(r){6-7}
    Method    & {\ab{stf}} & {\ab{ltd}} & {\ab{stf}} & {\ab{ltd}} & {\ab{stf}} & {\ab{ltd}} \\
    \midrule
    \mw{}     &  0.55 & -0.04 &  0.55 & -0.03 &  0.57 & -0.03 \\
    \Gext{3}  &  0.55 & -0.42 &  0.57 & -0.15 &  0.53 & -0.04 \\
    \Gext{6}  &  0.56 & -0.53 &  0.52 & -0.13 &  0.57 & -0.04 \\
    \bottomrule
  \end{tabular}
\end{table}

\section{Conclusion}\label{sec:conclusion}

In this contribution, we presented an extrapolation scheme to predict initial guesses of the
density matrix for the \ab{scf}-iterations within \ab{bomd}.
What makes our approach new is that we enforce the idempotency of the density matrix by
extrapolating not the densities themselves, but their map onto a vector space, which is the
tangent plane to the manifold of the physically acceptable densities.
Such a map is locally bijective, so that after performing the extrapolation, we can map the
new density back to the original manifold, providing thus an idempotent density.
The main element of novelty of the algorithm is that, by working on a tangent space, it
allows one to use any linear extrapolation technique, while at the same time automatically
ensuring the correct geometrical structure of the density matrix.
As such, the technique presented in this paper can be seen as a simple case of a more
general framework.
Such a framework allows one to recast the problem of predicting a guess density by
extrapolating information available from previous \ab{md} steps as a mapping between two
vector spaces, \latin{i.e.}, the space of molecular descriptors and the tangent plane.
This geometric approach can be seen as an alternative to extrapolating quantities derived
from the density, such as the Fock or Kohn--Sham matrix, as proposed by Pulay and
Fogarasi~\cite{pulay_fock_2004} and by Herbert and
Head-Gordon~\cite{herbert_accelerated_2005}.
However, the framework we developed, using molecular descriptors and a general linear
extrapolation technique, can in principle be easily extended to such approaches.

That being said, our choices of both the molecular descriptor and of the extrapolation
strategies are far from being unique.
In recent years, molecular descriptors gained attraction within the rise of \ab{ml}
techniques.
Our choice, namely, using the Coulomb matrix, is only one of the many possibilities, and
while being simple and effective, more advanced descriptors may be used and possibly improve
the overall performances of the method.
We also used a straightforward (stabilized) least-square interpolation of the descriptors at
previous point to compute the extrapolation coefficients for the densities.
This strategy is, again, simple yet effective.
However, many other approaches can be used.
In particular, \ab{ml} techniques may not only provide a very accurate approximated map, but
also benefit of a larger amount of information (\latin{i.e.}, use the densities computed at
a large number of previous steps), further improving the accuracy of the guess.
Improvements on the descriptors and extrapolation strategies are not the only possible
extensions of the proposed method.
A natural extension that is under active investigation is the application to the \ab{gext}
guess to geometry optimization, for which the \ab{xlbo} scheme cannot be used.

Overall, even the simple choices made in this contribution produced an algorithm that
exhibits promising performances.
In all our tests, the \ab{gext} method outperformed the well-established \ab{xlbo}
technique, especially for tighter \ab{scf} accuracies which may be relevant for
post-\ab{scf} \ab{bomd} computations, including computations on excited-state \ab{pes}.
While we tested the method only for \ab{ks} \ab{dft}, it can also be used for Hartree--Fock
or semiempirical calculations.
The main disadvantage of the proposed strategy with respect to the \ab{xlbo} method is,
however, the lack of time reversibility, which manifests itself as a lack of long-term
energy conservation.
In particular, for longer \ab{md} simulations, the total energy may exhibit a visible drift,
which is something that the user must be aware of.
In our test, the observed drift was relatively small and the use of a thermostat should be
enough to avoid problems in practical cases, however, this is a clear, and expected,
limitation of the proposed approach.
We note that, using a tighter \ab{scf} convergence, which is also the case where the
proposed method shows its best performances, produces an energy conserving trajectory, even
starting from very noisy conditions.
A time-reversible generalization of the \ab{gext} method is anyways particularly attractive,
and is at the moment under active investigation.

\begin{suppinfo}
  A \lang{Julia} template of the \ab{gext} algorithm is available at
  \url{https://github.com/epolack/GExt.jl}.
  The figure representing the total energy computation with an \ab{scf} convergence
  threshold of \num{e-6} for the molecule \ab{thf} and formulas for the exponential and
  logarithm functions are available in the supplementary information.
\end{suppinfo}

\section*{Funding}

Part of this work was supported by the French ``Investissements d’Avenir'' program, project
ISITE-BFC (contract ANR-15-IDEX-0003).

\bibliography{biblio}

\newpage

\includepdf[pages=-]{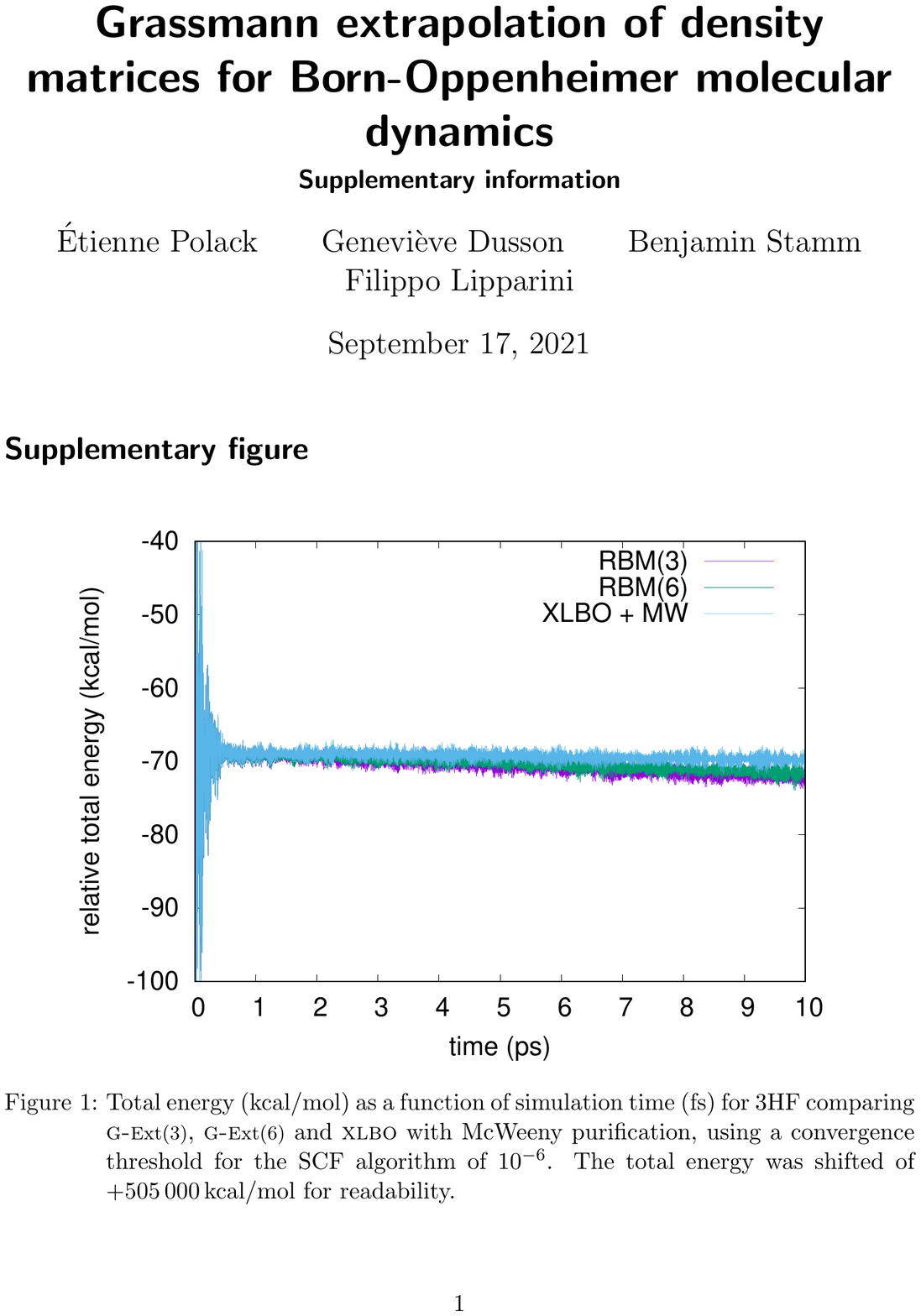}

\end{document}